\documentclass[twocolumn,english,journal, final]{IEEEtran}
\usepackage[T1]{fontenc}
\usepackage{geometry}
\usepackage{units}
\usepackage{amsmath}
\usepackage{amssymb}
\usepackage{graphicx}

\makeatletter

\providecommand{\tabularnewline}{\\}

\usepackage{babel}
\usepackage[caption=false,font=footnotesize]{subfig}

\usepackage{tikz}
\usetikzlibrary{dsp,chains}
\usepackage{pgfplots}
\usepackage{cite}

\addto\captionsenglish{}

\@ifundefined{showcaptionsetup}{}{%
 \PassOptionsToPackage{caption=false}{subfig}}
\usepackage{subfig}
\makeatother

\usepackage{babel}
\begin{document}

\title{SINR Analysis of Different Multicarrier Waveforms over Doubly Dispersive
Channels }

\author{Xiaojie~Wang,~\IEEEmembership{Student~Member,~IEEE,}
and~Stephan~ten~Brink,~\IEEEmembership{Senior~Member,~IEEE}
\thanks{The authors are with the Institute of Telecommunications, Pfaffenwaldring 47, University of Stuttgart, 70569 Stuttgart, Germany (e-mail: \{wang, tenbrink\}@inue.uni-stuttgart.de).}}
\maketitle
\begin{abstract}
Wireless channels generally exhibit dispersion in both time and frequency
domain, known as doubly selective or doubly dispersive channels. To
combat the delay spread effect, multicarrier modulation (MCM) such
as orthogonal frequency division multiplexing (OFDM) and its universal
filtered variant (UF-OFDM) is employed, leading to the simple per-subcarrier
one tap equalization. The time-varying nature of the channel, in particular,
the intra-multicarrier-symbol channel variation induces spectral broadening
and thus inter-carrier interference (ICI). Existing works address
both effects separately, focus on the one effect while ignoring the respective other.
This paper considers both effect simultaneously for cyclic prefix (CP)-, zero padded (ZP)- and UF-based
OFDM with simple one tap equalization, assuming a general wireless channel
model. For this general channel model, we show that the independent (wide
sense stationary uncorrelated scatter, WSSUS) selectivity
in time and frequency starts to intertwine in contrast to the ideal
cases with single selectivity. We derive signal-to-interference-plus-noise
ratio (SINR) in closed form for arbitrary system settings and channel
parameters, e.g., bandwidth, delay- and Doppler-spread. With the SINR
analysis, we compare the three MCM schemes under different channel
scenarios. 
\end{abstract}

\section{Introduction}

Multicarrier modulation (MCM), particularly orthogonal frequency division
multiplexing (OFDM), is proven to be the most practical technique
in terms of spectral efficiency and transceiver complexity. Relying
on orthogonal subcarriers with lower data rate, the channel-induced
inter-symbol interference (ISI) is significantly suppressed in comparison
to single carrier systems. With the help of a cyclic prefix (CP) and/or zero
padding (ZP) \cite{MuquetZPTC02}, though sacrificing some spectral efficiency,
ISI can be completely mitigated in OFDM so that the equalization procedure
is drastically simplified to a single tap. 

The orthogonality between OFDM subcarriers is often questioned in practical
systems because of nonlinear impairments of RF components such as
power amplifier and inaccurate transmit and receive oscillators. This is  even
more critical in OFDMA scenarios, i.e., multi-user uplink transmissions, where each user has its own RF chain
making the maintenance of orthogonality much more demanding. To improve
the spectral properties of OFDM, thus reduce inter-carrier interference
(ICI) in the absence of orthogonality, a subband-based filtering approach
was proposed and studied in \cite{UFMCtenBrink,WildUFOFDM,FrankISWCS14UFMCSyn,TCOM17Wang,TWC17Zhang,AdvancedUFMCTVT18,CST16Farhang,SPL16Zhang,TCOM17ZhangZ,Access17Liu,TWC18Chen,JSAC17Han,TVT18Han},
termed universal filtered OFDM (UF-OFDM). The redundant cyclic
prefix in OFDM, which serves as guard time between symbols, is
devoted to FIR-filtering \cite{XWangUFMCFilterOpt,FilterOptUFMCICCW16,Access18Wen}
of a subband in UF-OFDM as the consequence of the Balian\textendash Low
theorem prohibiting good per-subcarrier spectral shaping with short
filter (comparable to CP/ZP, e.g., $8\%-15\%$ of symbol length). 

The trend of future communication systems is to be more heterogeneous,
support faster vehicles and operate at higher carrier frequencies.
The often ignored intra OFDM symbol channel variation, i.e., time
selectivity, starts to play a role. 
In \cite{DopplerICIBoundsLi}\cite{DopplerRobertsonVTCFall1999},
the effect and/or bound of Doppler spread with Jakes, uniform and
two path model is analyzed. The equalization of Doppler spread in
the context of OFDM is studied in \cite{PSchniterTSPDDEqualization}\cite{PSchniterTSPDDmaxSINR}
and \cite{OFDMCommTVBookChapter}. However, all the existing literature
assume CP/ZP length larger than the maximum channel delay spread so
that frequency selectivity has no impact on the Doppler-induced ICI.
As pointed out in \cite{BEROFDMDopplerWang}, frequency-selectivity
of the channel has only moderate impact on the performance. Therefore, it is often
neglected in the vast majority of works. To the best of the authors'
knowledge, a comprehensive study of ICI and inter-symbol interference
(ISI) in general cases of MCM parameters and channel statistics has
not yet been done. Although the channel fading in time and frequency
is believed to be independent of each other, the interplay complicates
the ICI and ISI analysis if the channel delay spread is not restricted
to within CP/ZP duration.

This paper addresses both time- and frequency-selectivity of general
wireless channels. Both ICI and ISI are analytically computed using
derived closed-form formulas assuming the low-complexity single tap
equalization. Both, time and frequency correlation functions,
are incorporated in the analytical formula, covering all cases of
CP/ZP lengths, channel delay spreads and Doppler spreads. We compare
the SINR performance of the classic CP- and ZP-OFDM with the novel
UF-OFDM in both uplink (multiuser with diverse selectivity parameters)
and downlink (multiuser with same selectivity parameters) settings.
The analysis reveals that CP- and ZP-OFDM is robust to delay spread
and UF-OFDM is robust to Doppler spread, which provides guideline
for waveform choice depending on application-constrained channel statistics.

\textit{Notation: }We use $\left(\cdot\right)^{H}$ to denote Hermitian
transpose, $\mathrm{diag}\left(\cdot\right)$ diagonal elements vector
of matrices and/or diagonal matrices of vectors, $\mathrm{tr}\left(\cdot\right)$
the trace operator. The modulo-$N$ operation is denoted by $\left\langle \cdot\right\rangle _{N}$
and the element in the $i$th row and $j$th column of the matrix
$\mathbf{A}$ is denoted by $\left[\mathbf{A}\right]_{ij}$. Occasionally,
we use $\mathbf{A}_{j}$ to denote the $j$th row vector of the matrix
$\mathbf{A}$ for brevity. Finally, $K$-point IDFT and DFT matrices
are denoted by $\mathbf{F}_{K}^{H}$ and $\mathbf{F}_{K}$ with normalized
power, respectively.

\section{System model and doubly selective channel}

A system with a bandwidth of $N$ subcarriers is considered. A user
moving at the speed of $v_{u}$ is allocated with $M$ consecutive
subbands for data transmission, each consisting of $N_{\mathrm{RB}}$
subcarriers. The $m$th transmit signal $\mathbf{x}_{m}$ of the user
using OFDM is expressed as 
\begin{equation}
\mathbf{x}_{m,\mathrm{O}}=\mathbf{C}_{\mathrm{A}}\cdot{\displaystyle \sum_{i=1}^{M}}\mathbf{F}_{i}^{H}\cdot\mathbf{X}_{m,i},\label{eq:CPOFDMsym}
\end{equation}
where $\mathbf{X}_{m,i}$ with the dimension $N_{\mathrm{RB}}\times1$
denotes the $m$th QAM symbol vector of the $i$th subband, $\mathbf{F}_{i}^{H}$
is the corresponding inverse Fourier transform matrix and $\mathbf{C}_{A}$
adds $L_{\mathrm{CP}}$ samples of cyclic (or zero) prefix after the
Fourier transform.

UF-OFDM applies FIR-filtering after the Fourier transform to improve
the spectral property instead of inserting cyclic prefix. To ensure
the same spectral efficiency, FIR filters with the order $L=L_{\mathrm{F}}=L_{\mathrm{CP}}$
are used. Furthermore, the filtering is performed in a subband basis,
i.e., the $m$th transmit symbol reads 
\begin{equation}
\mathbf{x}_{m,\mathrm{U}}={\displaystyle \sum_{i=1}^{M}}\mathbf{G}_{i}\cdot\mathbf{F}_{i}^{H}\cdot\mathbf{X}_{m,i},\label{eq:UFOFDMsym}
\end{equation}
where $\mathbf{G}_{i}$ is a Toeplitz matrix comprising of the subband
filter coefficients $[g_{i,0},\cdots,g_{i,L_{\mathrm{F}}}]$. The
same type of filter with adjusted center frequency is often used for
each subband.

The wireless channel is in general time and frequency selective, which
results from dispersion (or broadening) in both time (multipath) and
frequency (Doppler). To account for both effects, the doubly selective
channel is often mathematically modeled by 
\begin{equation}
\underset{\left[\left(N+L+D\right)\,\times\,\left(N+L\right)\right]}{\mathbf{h}_{m}}=\left[\begin{array}{ccccc}
h_{00} &  & 0 &  & 0\\
h_{10} & h_{01}\\
\vdots & h_{11} & \ddots\\
h_{D0} & \vdots & \ddots &  & h_{0N+L}\\
 & h_{D1} &  &  & h_{1N+L}\\
0 & 0 & \ddots &  & \vdots
\end{array}\right],\label{eq:DDChannelMat}
\end{equation}
where $h_{ij}\thicksim\mathcal{CN}\left(0,\rho_{i}\right)$ is circularly
symmetric complex Gaussian distributed with zero mean and variance
$\rho_{i}$. It denotes the channel response of the $i$th tap at
the $j$th time instance and $D$ is the memory order of the channel.
Under the wide sense stationary uncorrelated scatterers (WSSUS) assumption,
it holds 
\begin{equation}
\mathrm{E}\left[h_{ij}\cdot h_{mn}^{*}\right]=\begin{cases}
0 & i\ne m\\
\rho_{i}R_{t}\left(\Delta n=j-n\right) & i=m
\end{cases},\label{eq:DDChannelTimeCorr}
\end{equation}
where $R_{t}\left(\Delta n\right)$ is the correlation between the
channel coefficients at two different time instances. In classic Clark
and Jakes model, the correlation function in time is given by 
\begin{equation}
R_{t}\left(\Delta n\right)=J_{0}\left(2\pi f_{D}T_{s}\cdot\Delta n\right),\label{eq:ClarkJakeCorrTime}
\end{equation}
where $J_{0}\left(\cdot\right)$ is the zero-th order Bessel function
of the first kind, $f_{D}$ denotes the Doppler frequency and $T_{s}$
denotes the time between two consecutive samples. Taking the Fourier
transform of the correlation function which results in the famous
Bathtub spectrum, the spectral broadening becomes obvious inducing
ICI. The multipath propagation on the other hand causes ISI, if not
properly dealt with, and frequency selectivity. An exponential power
delay profile is investigated throughout this paper, i.e., $\rho_{i}=\rho_{0}\beta^{i}$,
if not otherwise stated. Briefly speaking, the power of the channel
impulse response decays exponentially with its delay. The frequency
correlation function can be obtained by taking Fourier transform of
the power delay profile (PDP), i.e., 
\begin{equation}
R_{f}\left(\Delta k\right)=\frac{\rho_{0}}{\sqrt{N}}\cdot\frac{1-\beta^{N}e^{j2\pi\Delta k}}{1-\beta e^{j2\pi\frac{\Delta k}{N}}}.\label{eq:EDPDPFreqCorr}
\end{equation}
Furthermore, the mean delay spread and root mean square (rms) delay
spread are given by the first moment and the standard deviation of
the PDP.

\section{SINR analysis}

For simplicity of notation, yet without loss of generalization, single
subband allocation is considered (i.e., $M=1$) in the following.
Let $\mathbf{x}_{m}$ denote either $\mathbf{x}_{m,\mathrm{O}}$ or
$\mathbf{x}_{m,\mathrm{U}}$ with a little abuse of notation. Consider
the detection of the $m$th (OFDM and/or UF-OFDM) symbol, the received
signal within the detection window consists of part of the signal
of interest and part of previous symbol due to the channel delay spread
with the assumption $D\leq N-L$ (so that merely the previous symbol
contributes to ISI), which reads 
\begin{equation}
\mathbf{y}_{m}=\mathbf{w}_{\mathrm{D}}\mathbf{h}_{m}\mathbf{x}_{m}+\mathbf{w}_{\mathrm{I}}\mathbf{h}_{m-1}\mathbf{x}_{m-1}+\mathbf{n}\label{eq:TDsigVec}
\end{equation}
where $\mathbf{n}$ denotes uncorrelated Gaussian noise with $\sigma_{n}^{2}\mathbf{I}$
auto-covariance, $\mathbf{w}_{\mathrm{D}}$ and $\mathbf{w}_{\mathrm{I}}$
are to model the detection window and ISI, they can be expressed by
\begin{align}
\mathbf{w}_{\mathrm{D}} & =\left[\begin{array}{cc}
\mathrm{diag\left(\mathbf{v}\right)} & \mathbf{0}_{\left(N+L\right)\times D}\end{array}\right]\label{eq:DetectionWindow}\\
\mathbf{w}_{\mathrm{I}} & =\left[\begin{array}{cc}
\mathbf{0}_{D\times\left(N+L\right)} & \mathrm{diag\left(\tilde{\mathbf{v}}\right)}\\
\mathbf{0}_{\left(N+L-D\right)\times\left(N+L\right)} & \mathbf{0}_{\left(N+L-D\right)\times D}
\end{array}\right],
\end{align}
where $\mathbf{v}=\left[v_{1},\cdots,v_{N+L}\right]$ contains the
coefficients of receive windowing and $\tilde{\mathbf{v}}=\left[v_{1},\cdots v_{D}\right]$
denotes the ISI partial window. To facilitate the following interference
analysis, we re-write \eqref{eq:TDsigVec} as follows 
\begin{equation}
\mathbf{y}_{m}=\mathbf{\widetilde{w}}_{\mathrm{D}}\tilde{\mathbf{h}}_{m}\tilde{\mathbf{x}}_{m}+\tilde{\mathbf{w}}_{\mathrm{I}}\tilde{\mathbf{h}}_{m-1}\tilde{\mathbf{x}}_{m-1}+\mathbf{n},\label{eq:TDSigVec2}
\end{equation}
where $\tilde{\mathbf{x}}_{m}$ and $\tilde{\mathbf{x}}_{m-1}$ with
the dimension $2N\times1$ are zero-padded versions of $\mathbf{x}_{m}$
and $\mathbf{x}_{m-1}$, respectively. $\tilde{\mathbf{h}}_{m}$ and
$\tilde{\mathbf{h}}_{m-1}$ are the time-variant, circular convolution
channel matrix with the dimension $2N\times2N$. Furthermore, the
window matrices are correspondingly padded with zero columns, resulting
in two $\left(N+L\right)\times2N$ matrices, $\mathbf{\widetilde{w}}_{\mathrm{D}}$
and $\mathbf{\widetilde{w}}_{\mathrm{I}}$.

In CP/ZP-OFDM systems, the cyclic/zero prefix is subsequently removed
and $N$-FFT is carried out. Hence, the frequency domain signal is
obtained as 
\begin{equation}
\begin{split}\mathbf{Y}_{m,\mathrm{O}} & =\mathbf{F}_{N}\mathbf{C}_{\mathrm{R}}\mathbf{y}_{m,\mathrm{O}}\\
= & \underset{\mathbf{Y}_{\mathrm{s,O}}+\mathbf{Y}_{\mathrm{ICI,O}}}{\underbrace{\mathbf{F}_{N}\mathbf{C}_{\mathrm{R}}\tilde{\mathbf{w}}_{\mathrm{D}}\tilde{\mathbf{h}}_{m}\tilde{\mathbf{x}}_{m,\mathrm{O}}}}+\underset{\mathbf{Y}_{\mathrm{ISI,O}}}{\underbrace{\mathbf{F}_{N}\mathbf{C}_{\mathrm{R}}\tilde{\mathbf{w}}_{\mathrm{I}}\tilde{\mathbf{h}}_{m-1}\mathbf{\tilde{x}}_{m-1,\mathrm{O}}}}+\mathbf{n},
\end{split}
\label{eq:OFDMRxSymFd}
\end{equation}
where $\mathbf{F}_{N}$ is the $N$-FFT matrix and $\mathbf{C}_{\mathrm{R}}=\left[\mathbf{0}_{N\times L}\,\mathbf{I}_{N}\right]$
models the CP/ZP removal. Since the statistical property of the noise
remains the same after Fourier transform, we still use $\mathbf{n}$
denoting the noise in the frequency domain. The first term can be
decomposed into the signal term $\mathbf{Y}_{\mathrm{s}}$ and the
ICI term $\mathbf{Y}_{\mathrm{ICI}}$ due to Doppler spread. The second
term is ISI due to delay spread of the channel. Provided that the
channel delay spread is within the CP/ZP duration, i.e., $D\leq L$,
this term vanishes, i.e., $\mathbf{Y}_{\mathrm{ISI}}=\mathbf{0}$.
For the more general case, i.e., $D\leq N-L$ (further generalization
is straightforward), it can be shown in Appendix \ref{A} that 
\begin{equation}
\mathbf{Y}_{\mathrm{ISI,O}}=\begin{cases}
\mathbf{0} & D\leq L\\
\mathbf{W}_{\mathrm{I}}\mathbf{H}_{m-1}\mathbf{T}_{\mathrm{O}}\mathbf{X}_{m-1} & L<D\leq N-L
\end{cases}.\label{eq:YISIOFDM}
\end{equation}
The expected ISI power can be thus computed by 
\begin{equation}
\mathbf{P}_{\mathrm{ISI,O}}=\mathrm{diag}\left(\mathbf{W}_{\mathrm{I}}\mathbf{R}_{\mathrm{H}}\mathbf{W}_{\mathrm{I}}^{H}\right).\label{eq:ISIOFDM}
\end{equation}
The detailed derivation is given in Appendix \ref{B}. We note that
our EDPDP channel model exhibits always the maximum delay spread of
$D_{\mathrm{max}}=N-L$, while its mean delay spread and rms delay
spread are parameterized by the decay factor $\beta$. The ISI power
therefore asymptotically approaches zero as $\beta$ approaches 0.
Similarly, the signal power and ICI power can be computed according
to 
\begin{align}
\mathbf{P}_{\mathrm{S,O}}=\mathrm{diag}\left(\mathbf{W}_{\mathrm{D}}\mathbf{R}_{\mathrm{H,S}}\mathbf{W}_{\mathrm{D}}^{H}\right)\label{eq:ICIAndSignal}\\
\mathbf{P}_{\mathrm{ICI,O}}=\mathrm{diag}\left(\mathbf{W}_{\mathrm{D}}\mathbf{R}_{\mathrm{H,ICI}}\mathbf{W}_{\mathrm{D}}^{H}\right)
\end{align}

UF-OFDM system differs from CP- and ZP-OFDM system in the following
aspects in view of signal processing. 1) The insertion of cyclic (or
zero) prefix is replaced by FIR-filtering, i.e., $\mathbf{C}_{\mathrm{A}}\rightarrow\mathbf{G}_{i}$;
2) The removal of CP/ZP is absent. 2) $2N$-FFT of $N+L$ samples
with subsequent downsampling rate of $2$ is performed, represented
by the $N\times\left(N+L\right)$ Fourier transform matrix $\mathbf{\tilde{F}}_{2N}$.
The frequency domain signal in UF-OFDM (again single subband analysis
suffices) is then given by 
\begin{align}
\mathbf{Y}_{m,\mathrm{U}} & =\mathbf{\tilde{F}}_{2N}\mathbf{y}_{m,\mathrm{U}}\nonumber \\
= & \underset{\mathbf{Y}_{\mathrm{s,U}}+\mathbf{Y}_{\mathrm{ICI,U}}}{\underbrace{\mathbf{\tilde{F}}_{2N}\tilde{\mathbf{w}}_{\mathrm{D}}\tilde{\mathbf{h}}_{m}\tilde{\mathbf{x}}_{m,\mathrm{U}}}}+\underset{\mathbf{Y}_{\mathrm{ISI},\mathrm{U}}}{\underbrace{\tilde{\mathbf{F}}_{2N}\tilde{\mathbf{w}}_{\mathrm{I}}\tilde{\mathbf{h}}_{m-1}\mathbf{\tilde{x}}_{m-1,\mathrm{U}}}}+\mathbf{n}.\label{eq:UFMCRxSymFd}
\end{align}
We show in Appendix \ref{sec:ISI-in-UF-OFDM} the ISI analysis of
UF-OFDM signal and omit the similar ICI and signal derivation. Following
the same analysis as CP/ZP-OFDM, we obtain 
\begin{align*}
\mathbf{P}_{\mathrm{S,U}}=\mathrm{diag}\left(\tilde{\mathbf{W}}_{\mathrm{D}}\tilde{\mathbf{R}}_{\mathrm{H,S}}\tilde{\mathbf{W}}_{\mathrm{D}}^{H}\right)\\
\mathbf{P}_{\mathrm{ICI,U}}=\mathrm{diag}\left(\mathbf{\tilde{W}}_{\mathrm{D}}\tilde{\mathbf{R}}_{\mathrm{H,ICI}}\tilde{\mathbf{W}}_{\mathrm{D}}^{H}\right)\\
\mathbf{P}_{\mathrm{ISI,U}}=\mathrm{diag}\left(\mathbf{\tilde{W}}_{\mathrm{I}}\mathbf{\tilde{R}}_{\mathrm{H}}\mathbf{\tilde{W}}_{\mathrm{I}}^{H}\right).
\end{align*}

\section{\label{sec:Simulation-Results}Results }

For the following analysis and performance comparison, we set the
total subcarrier number to $N=1024$, CP/ZP length $73$ and FIR-filter
length $74$ (Chebyshev filter with sidelobe attenuation of $\unit[40]{dB}$).
The FIR filtering is performed in a subband manner each with $N_{\mathrm{RB}}=12$
subcarriers. QAM-symbols modulated at each subcarriers are assumed
to be uncorrelated and with unit variance. The receiver employs rectangular
window and the noise floor is set to $\unit[-40]{dB}$. We investigate
the channel PDP with exponentially decaying power, yet unit power,
as aforementioned and the delay stretch between two taps is $8$ samples.
The time variation of the channel is according to the classic Jakes
and Clark's model.

\subsection{Verification}

The signal, ICI and ISI power analysis based on the channel PDP and
Doppler spectrum (or time correlation model) are compared with Monte-Carlo-based
simulations, each averaged over $10^{4}$ channel realizations, in
Fig. \ref{fig:Signal,-ICI-and}. For this verification, the channel
PDP is according to the standardized Vehicular-B model in which the
maximum delay spread is larger than the CP/ZP length in the considered
OFDM. Furthermore, the rms delay spread equals $\tau_{\mathrm{rms}}T_{s}^{-1}=61.6$
and one single subband with QPSK signaling is considered. 
\begin{figure}[tbh]
\begin{centering}
\subfloat[VehB-Channel \cite{VEHAB} with $f_{\mathrm{D}}T_{s}=3\times10^{-5}$\label{fig:UF-OFDM:-VehB-Channel-with}]{\begin{centering}
\includegraphics[width=1\columnwidth]{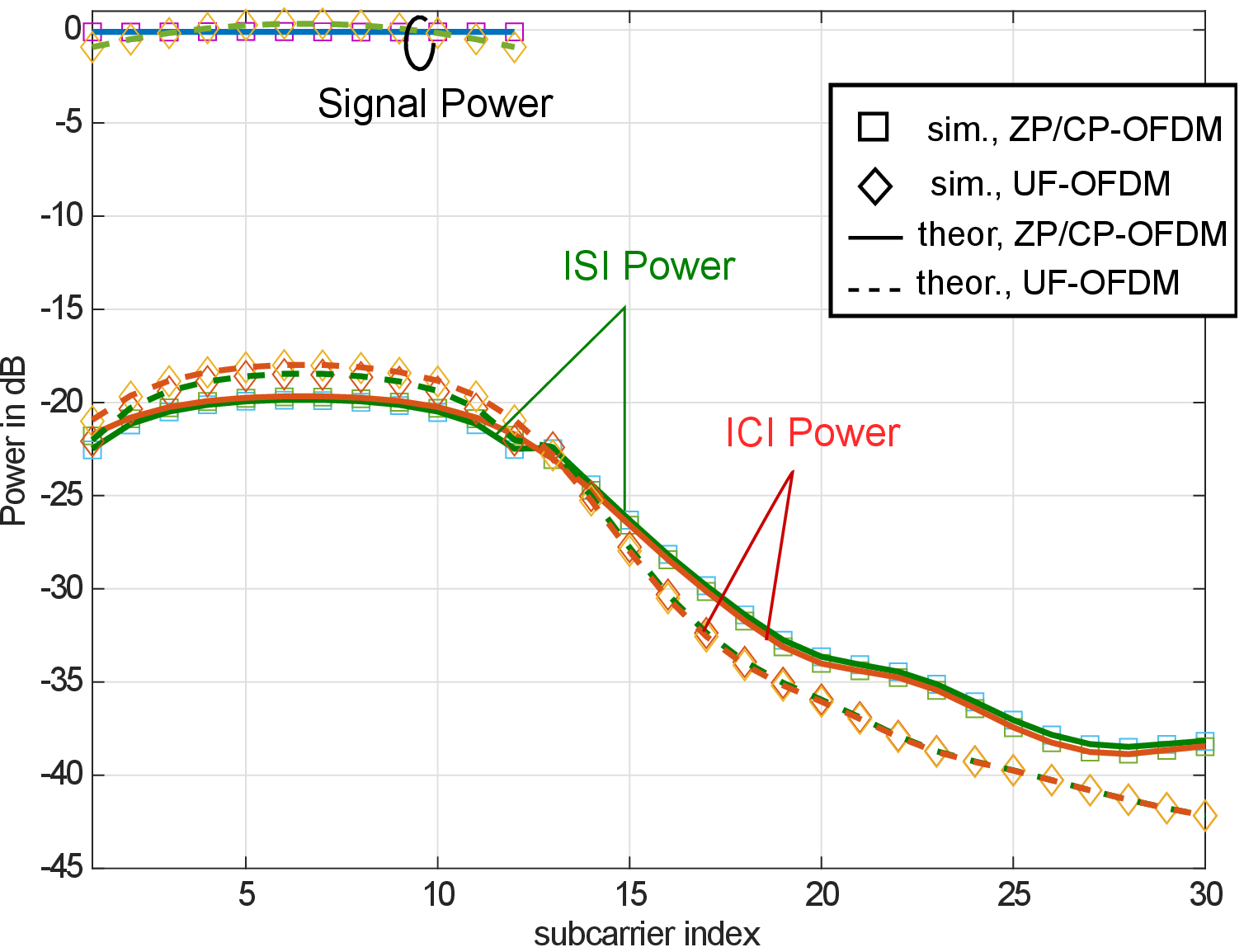}
\par\end{centering}
}
\par\end{centering}
\subfloat[VehB-Channel \cite{VEHAB} with $f_{\mathrm{D}}T_{s}=1.5\times10^{-3}$\label{fig:CP-OFDM:-VehB-Channel-with}]{\begin{centering}
\includegraphics[width=1\columnwidth]{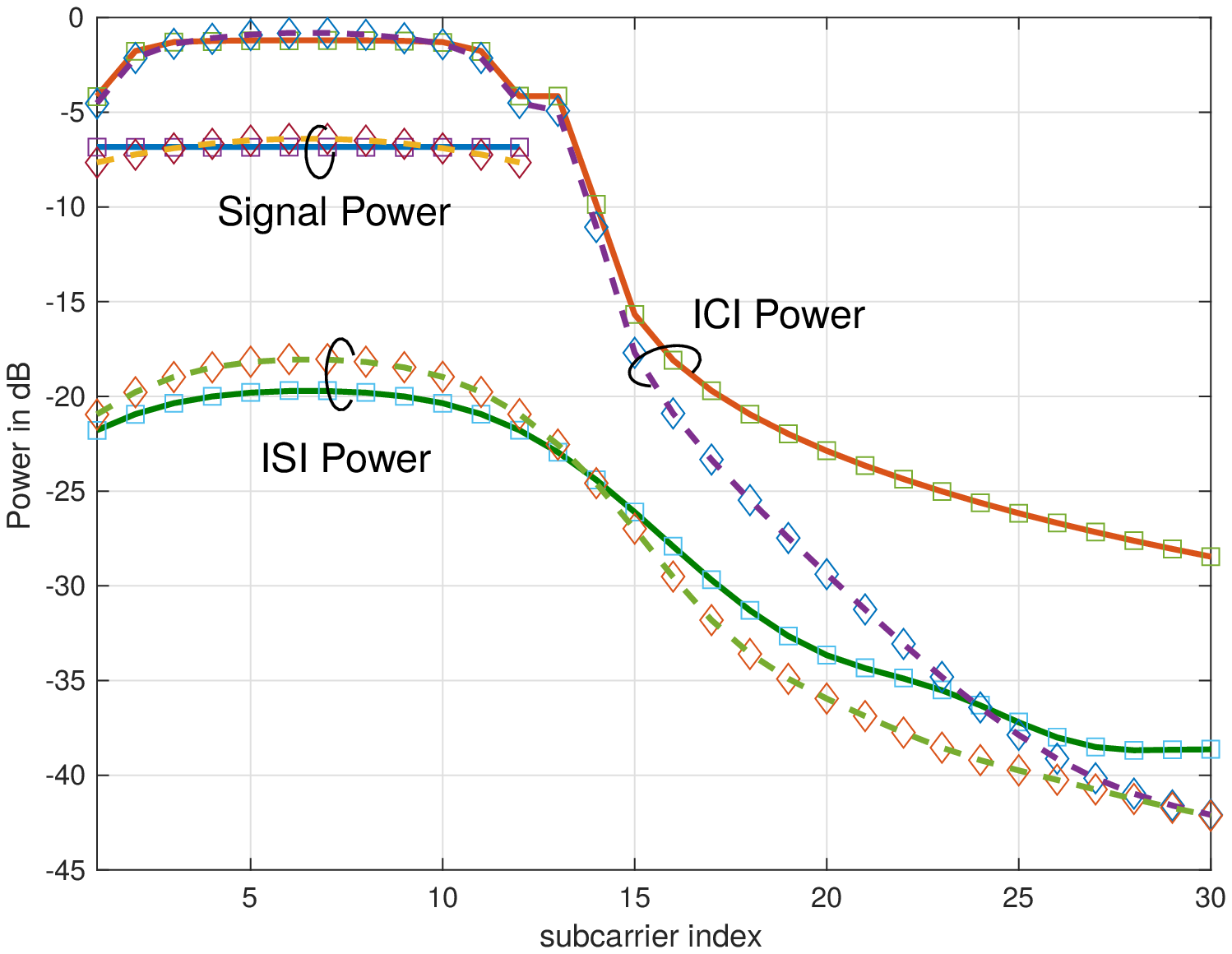}
\par\end{centering}

}

\caption{Signal, ICI and ISI power analysis; simulation versus theoretical
analysis\label{fig:Signal,-ICI-and}}

\end{figure}
In Fig. \ref{fig:UF-OFDM:-VehB-Channel-with}, the power distributions
of the signal, ICI and ISI component are plotted over subcarriers
with UF-OFDM and CP/ZP-OFDM waveforms respectively where a relatively
small Doppler spread of $f_{\mathrm{D}}T_{s}=3\times10^{-5}$ is considered.
The results obtained by simulations and the derived analytical approach
match very well. The signal power variation is due to the FIR-filtering
in UF-OFDM. For comparably large Doppler spread of $f_{\mathrm{D}}T_{s}=1.5\times10^{-3}$,
Fig. \ref{fig:CP-OFDM:-VehB-Channel-with} shows the signal, ICI and
ISI power level respectively. In contrast to UF-OFDM, the signal power
of CP/ZP-OFDM remains constant over subcarriers. The large signal
power degradation is mainly due to the large Doppler spread. While
the ICI level increases because of Doppler spread, the ISI power hardly
changes with Doppler spreads (It means that the average ISI power
over the entire bandwidth is independent of the Doppler spread while
its distribution over subcarriers slightly changes). We note that
the Monte-Carlo simulation approach is quite computational intensive
because of the correlation between the channel coefficients at arbitrary
two time instants in comparison to the analytical approach relying
solely on the correlation statistics in time and frequency domain.
It is also noteworthy to mention that CP/ZP-OFDM is capable of mitigating
delay spread induced ISI and ICI while UF-OFDM generates less out-of-band
emission to adjacent channels/subcarriers.

\subsection{Downlink SINR}

\begin{figure}[tbh]
\begin{centering}
\includegraphics[width=1\columnwidth]{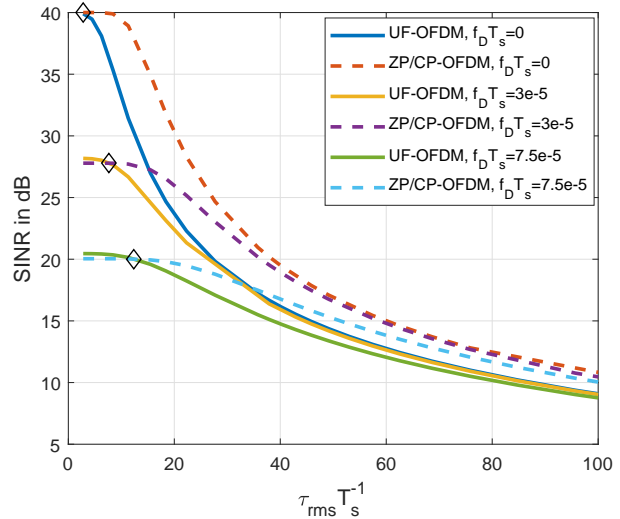}
\par\end{centering}
\caption{SINR versus delay spread\label{fig:SINR-versus-delay}}
\end{figure}

In the downlink, $M=85$ subbands are occupied for data transmission
(to multiple and/or single user) which corresponds to $1020$ subcarriers.
The downlink signal arrives at one user through a wireless channel
with certain delay spread and Doppler spread. The average SINR over
all the subcarriers are shown in the following. Fig. \ref{fig:SINR-versus-delay}
shows the SINR performance 
of CP/ZP-OFDM and UF-OFDM over a variety of rms channel delay spread.
Due to the absence of CP/ZP in UF-OFDM, it is vulnerable to delay
spread whereas CP/ZP-OFDM achieves much high SINR thanks to CP/ZP.
With increasing Doppler spread, the SINR decreases because of ICI.
It can be observed that UF-OFDM slightly outperforms CP/ZP-OFDM at
the low delay spread region since the filtering in UF-OFDM improves
its spectral compactness leading to less inter-subband interference
compared to CP/ZP-OFDM. The intersection of two curves with the same
Doppler spread parameter, i.e., that of UF- and CP/ZP-OFDM respectively,
are marked by diamond in the figure. The SINR gain is not significant
due to two reasons; the FIR-filter is not optimized; the intra-subband
interference caused by Doppler spread is much larger.

\begin{figure}[tbh]
\begin{centering}
\includegraphics[width=1\columnwidth]{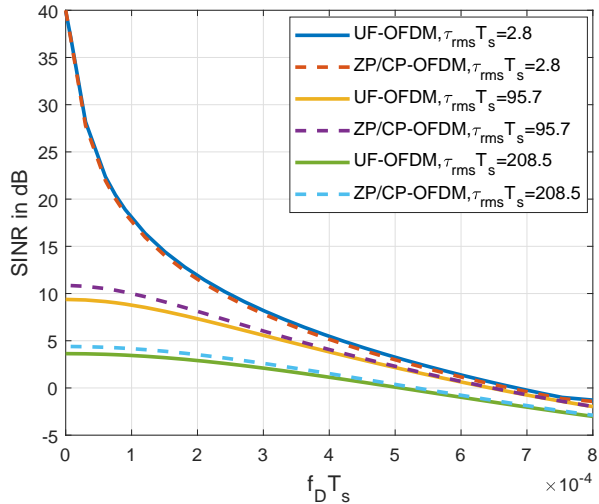}
\par\end{centering}
\caption{SINR versus Doppler spread\label{fig:SINR-versus-Doppler}}
\end{figure}

Next, we show the SINR performance of both system over Doppler spreads
in Fig. \ref{fig:SINR-versus-Doppler}. The inserted CP/ZP is extremely
effective for mitigating delay spread -induced ISI and ICI. It has
no impact on the Doppler-induced ICI. From the results, it can be
concluded that both waveform techniques have similar performance.
To be more precisely, UF-OFDM shall be favored with low delay spread
high Doppler spread channels and CP/ZP-OFDM is superior with large
delay spread and low Doppler spread channels.

\begin{figure}[tbh]
\begin{centering}
\includegraphics[width=1\columnwidth]{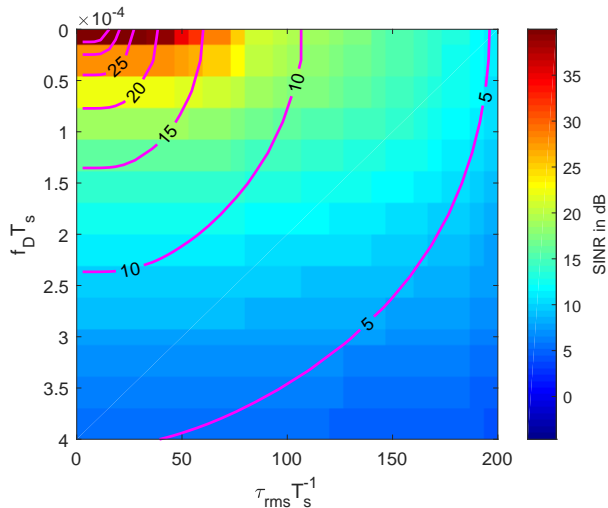}
\par\end{centering}
\caption{SINR of CP-OFDM\label{fig:SINRheatMapCPOFDM}}
\end{figure}

\begin{figure}[tbh]
\begin{centering}
\includegraphics[width=1\columnwidth]{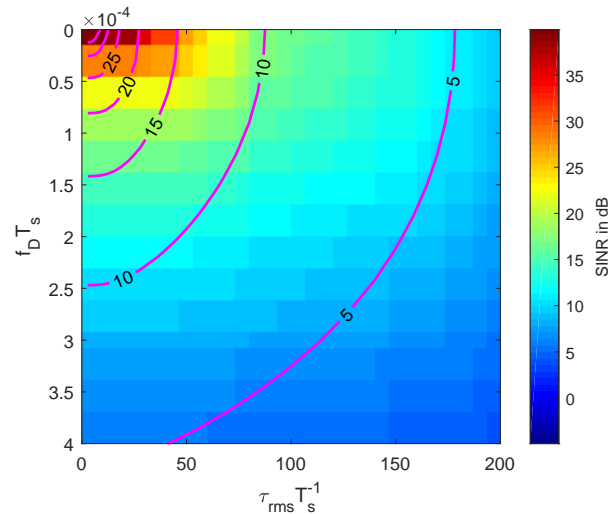}
\par\end{centering}
\caption{SINR of UF-OFDM\label{fig:SINRheatMapUFOFDM}}
\end{figure}

Finally, we show in Fig. \ref{fig:SINRheatMapCPOFDM} and in Fig.
\ref{fig:SINRheatMapUFOFDM} the SINR heat-map for different channels
with time and frequency selectivity parameters with CP/ZP-OFDM and
UF-OFDM waveforms, respectively. 
\begin{figure}[tbh]
\begin{centering}
\includegraphics[width=1\columnwidth]{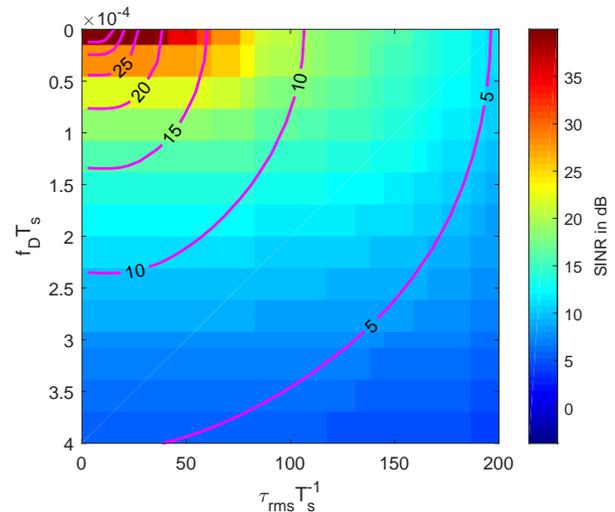}
\par\end{centering}
\caption{SINR of ZP-OFDM\label{fig:SINRheatMapCPOFDM-1}}
\end{figure}
The contour lines are also depicted in both figures. It is obvious
that CP/ZP-OFDM provides better delay spread protection than UF-OFDM.
It may not be so obvious from the figures that UF-OFDM provides small
to marginal gain over CP/ZP-OFDM with increasing Doppler-spreads.
It seems that in the downlink transmission, CP/ZP-OFDM outperforms
UF-OFDM in terms of delay spread while it achieves only slightly worse
performance at same channel Doppler-spread. However, we remark that
cyclic prefix or zero pre-/postfix can also be easily inserted in
UF-OFDM for delay spread protection with further loss of spectral
efficiency.

\subsection{Uplink SINR}

\begin{figure}[tbh]
\begin{centering}
\includegraphics[width=0.45\textwidth]{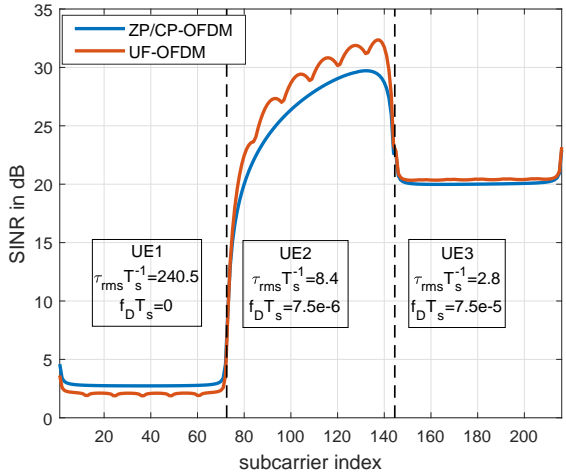}
\par\end{centering}
\caption{SINR over frequencies in a heterogeneous uplink multiuser scenario\label{fig:SINR-over-frequencies}}
\end{figure}

In the uplink, the signals of different users arrive at the basestation
through very different channels. The SINR is thus very different,
depending not exclusively on its own channel conditions but also on
the adjacent channel conditions. We consider three user equipments
(UEs) each with the bandwidth of $6$ subbands. The per-subcarrier
SINR is shown in Fig. \ref{fig:SINR-over-frequencies} along with
the channel conditions of each UE. 
The UE1 is stationary but with relatively long channel delay; the
UE2moves at relatively slow speed (corresponding to $\unit[50]{kmh}$
with the carrier frequency of $\unit[2.5]{GHz}$ and total bandwidth
of $\unit[15]{MHz}$) and the channel delay spread is also comparatively
small; the UE3 moves at very high speed and it experiences nearly
flat fading. UE1 achieves $\unit[0.7]{dB}$ SINR gain with CP/ZP-OFDM
while it generates larger interference to the adjacent UE2. UE3 achieves
$\unit[0.3]{dB}$ SINR gain with UF-OFDM and generates less interference
to UE2 simultaneously for the high Doppler low delay channel case.
In this heterogeneous traffic setting, the UE2 can achieve the SINR
gain of $\unit[2.1]{dB}$ with UF-OFDM. 
\begin{table}[tbh]
\renewcommand{\arraystretch}{1.5}
\begin{center}
\begin{tabular}{|c|c|c|}
\hline 
 & CP/ZP-OFDM & UF-OFDM\tabularnewline
\hline 
\hline 
UE1 & 1.55~bpcu & 1.40~bpcu\tabularnewline
\hline 
UE2 & 9.04~bpcu & 9.74~bpcu\tabularnewline
\hline 
UE3 & 6.73~bpcu & 6.85~bpcu\tabularnewline
\hline 
sum & 17.32~bpcu & 17.99~bpcu\tabularnewline
\hline 
\end{tabular}
\end{center}
\vspace*{10pt}
\caption{Channel Capacity\label{tab:Channel-Capacity}}
\end{table}
In Tab. \ref{tab:Channel-Capacity}, the (lower bound) of the channel
capacity per UE and the sum capacity is compared. The UE1 looses $\unit[0.15]{bpcu}$
(bit per channel use) applying filtering instead of CP/ZP, however,
by doing this it is capable of boosting the adjacent users' channel
capacity by $\unit[0.7]{bpcu}$ leading to the overall sum capacity
gain of $\unit[0.67]{bpcu}$.

\subsection{Discussion}

The results reveal that CP/ZP-OFDM is profoundly resilient to delay
spread and vulnerable to Doppler spread while UF-OFDM on the opposite
offers better Doppler-spread protection, both relying on the same
amount of redundancy in the form of filtering and CP/ZP. In the case
of downlink transmission where the channel remains homogeneous within
entire bandwidth, CP/ZP-OFDM can operate on a wider range of channel
characteristics while UF-OFDM (yet with non optimized filtering procedure)
offers slightly SINR gain in low delay spread region. However, if
channel characteristics of adjacent channels are extremely diverse
as in the uplink, UF-OFDM might be better to support a larger variety
of users with different speed profiles. Furthermore, we note that
the FIR-filter is somewhat arbitrarily chosen in UF-OFDM and not optimized
for the channel statistics. An optimal design of FIR-filtering (including
zero-padding or cyclic prefix) could be an interesting extension of
this work.

\section{Conclusion}

The impact of delay spread and Doppler spread of general wireless
channels is analytically addressed in the context of orthogonal multicarrier
waveforms, particularly CP/ZP-OFDM and UF-OFDM. The derived analytical
formula is in closed-form assuming known Doppler-spectrum, power delay
profile (PDP) and simple one-tap equalizer. The results can be applied
to any channel and ZP-/CP-/UF-OFDM transceiver parameters in terms
of delay, Doppler spread, ZP/CP/filter length and receive windows.
With the obtained analytical formula, the SINR analysis and comparison
between the waveforms is facilitated without requiring computationally
extensive simulations. The SINR comparison shows that CP/ZP-OFDM can
efficiently mitigate the ISI/ICI caused by channel delay spread and
is vulnerable to Doppler spread while UF-OFDM just exhibits the opposite
property. Future works involve the FIR-filter optimization and the
window design for both waveforms. More sophisticated and low complexity
equalizer design and optimal combination of FIR-filtering and CP/ZP-insertion
could be tackled.

\section*{Acknowledgment }

The authors are grateful to Ms. Anupama Hegde, Mr. Maximilian Arnold
and Mr. Alexander Knaub for their kind support and the fruitful discussions.

\appendices{}

\section{ISI in CP/ZP-OFDM\label{A}}

Inserting \eqref{eq:TDSigVec2} and \eqref{eq:CPOFDMsym} into \eqref{eq:OFDMRxSymFd},
we obtain 
\[
\mathbf{Y}_{\mathrm{ISI,O}}=\mathbf{F}_{N}\mathbf{C}_{\mathrm{R}}\tilde{\mathbf{w}}_{\mathrm{I}}\tilde{\mathbf{h}}_{m-1}\tilde{\mathbf{C}}_{\mathrm{A}}\mathbf{F}^{H}\cdot\mathbf{X}_{m-1}
\]
where $\tilde{\mathbf{C}}_{\mathrm{A}}$ is padded with zero rows
to align the matrix dimension to $2N\times N$. Next, we incorporate
the frequency and Doppler description of the time-varying channel
matrix into the equation, which reads 
\[
\mathbf{Y}_{\mathrm{ISI},\mathrm{O}}=\underset{\mathbf{W}_{\mathrm{I}}}{\underbrace{\mathbf{F}_{N}\mathbf{C}_{\mathrm{R}}\tilde{\mathbf{w}}_{\mathrm{I}}\mathbf{F}_{2N}^{H}}}\underset{\mathbf{H}_{m-1}}{\underbrace{\mathbf{F}_{2N}\tilde{\mathbf{h}}_{m-1}\mathbf{F}_{2N}^{H}}}\underset{\mathbf{T}_{\mathrm{O}}}{\underbrace{\mathbf{F}_{2N}\tilde{\mathbf{C}}_{\mathrm{A}}\mathbf{F}^{H}}}\cdot\mathbf{X}_{m-1}.
\]
If the CP/ZP length exceeds maximum delay spread, i.e., $D\leq L$,
then $\mathbf{C}_{\mathrm{R}}\tilde{\mathbf{w}}_{\mathrm{I}}=\mathbf{0=W}_{\mathrm{I}}$,
hence no ISI is present. More generally, $\mathbf{W}_{\mathrm{I}}$
is highly structured like circulant matrices. Let $\mathbf{\boldsymbol{\mathbf{\omega}}}_{i}$
be the $i$th row vector of $\mathbf{W}_{\mathrm{I}}$, it can be
obtained by cyclically shifting the first row by $2i$ elements and
multiplying a phase shift, i.e., $\boldsymbol{\omega}_{i}=\mathrm{circshift}\left[\boldsymbol{\omega}_{0}\cdot e^{j4\pi iL/N},2i\right].$
Its first row vector is simply the 2N-IFFT of the ISI window, i.e.,
$\boldsymbol{\omega}_{0}^{T}=\frac{1}{\sqrt{N}}\mathbf{F}_{2N}^{H}\cdot\left[\mathbf{0}_{N+2L,}\:v_{L+1},\cdots,v_{D},\mathbf{0}_{N-L-D}\right]^{T}.$
$\mathbf{H}_{m-1}$ is one realization of the 2 dimensional Fourier
transform of the doubly selective channel of the $m-1$th symbol.
$\mathbf{T}_{\mathrm{O}}$ models the pulse shaping operation at transmitter
(or CP/ZP insertion in OFDM), which has the similar structure as $\mathbf{W}_{\mathrm{I}}$,
i.e., $\boldsymbol{t}_{i}=\mathrm{circshift}\left[\boldsymbol{t}_{0}\cdot e^{-j2\pi iL/N},2i\right].$
Its $i$th column $\mathbf{t}_{i}$ is cyclically shifted of $\mathbf{t}_{1}$
by $2i$ and $\mathbf{t}_{0}=\frac{1}{\sqrt{N}}\mathbf{F}_{2N}\cdot\left[\mathbf{1}_{N+L},\mathbf{0}_{N-L}\right]^{T}$. 

\section{ISI in UF-OFDM\label{sec:ISI-in-UF-OFDM}}

Inserting \eqref{eq:UFOFDMsym} into \eqref{eq:UFMCRxSymFd}, we obtain
\[
\mathbf{Y}_{\mathrm{ISI,U}}=\underset{\mathbf{W}_{\mathrm{I,U}}}{\underbrace{\tilde{\mathbf{F}}_{2N}\tilde{\mathbf{w}}_{\mathrm{I}}\mathbf{F}_{2N}^{H}}}\underset{\mathbf{H}_{m-1}}{\underbrace{\mathbf{F}_{2N}\tilde{\mathbf{h}}_{m-1}\mathbf{F}_{2N}^{H}}}\underset{\mathbf{T}_{\mathrm{U}}}{\underbrace{\mathbf{F}_{2N}\tilde{\mathbf{G}}\mathbf{F}^{H}}}\cdot\mathbf{X}_{m-1},
\]
where $\mathbf{W}_{\mathrm{I,U}}$ has the same ``circulant'' property
as in CP/ZP-OFDM. However, its first row is given by $\boldsymbol{\omega}_{0,\mathrm{U}}^{T}=\frac{1}{\sqrt{N}}\mathbf{F}_{2N}^{H}\cdot\left[\mathbf{0}_{N+L,}\:v_{1},\cdots,v_{D},\mathbf{0}_{N-L-D}\right]^{T}$
and the per row constant phase shift equals $e^{j2\pi iL/N}$. This
leads to nonzero ISI whenever $D\geq0$ due to the absence of guard
interval in time. Because of FIR-filtering, the filter response is
incorporated in the pulse-shaping matrix, it can be show that $\mathbf{T}_{\mathrm{U}}=\mathrm{diag}\left(\mathbf{F}_{2N}[g_{i,0},\cdots,g_{i,L_{\mathrm{F}}},\mathbf{0}_{2N-L_{F}-1}]\right)\cdot\tilde{\mathbf{T}}_{\mathrm{U}}$
and $\tilde{\mathbf{T}}_{\mathrm{U}}$ has the same ``circulant''
property as $\mathbf{T}_{\mathrm{O}}$. The first column of $\tilde{\mathbf{T}}_{\mathrm{U}}$
is given by $\tilde{\mathbf{t}}_{1}=\sqrt{2}\mathbf{F}_{2N}\cdot\left[\mathbf{1}_{N},\mathbf{0}_{N}\right]^{T}$.

\section{ISI power\label{B}}

We derive the average ISI power per subcarrier for both UF- and CP/ZP-OFDM,
since the two systems differ only in pulse shaping and receive processing.
Consider the ISI part, 
\begin{equation}
\mathbf{Y}_{\mathrm{ISI}}=\mathbf{W}_{\mathrm{I}}\mathbf{H}_{m-1}\mathbf{T}\mathbf{s}_{m-1}
\end{equation}
where $\mathbf{T}$ denotes the pulse-shaping matrix for either UF-
or CP/ZP-OFDM signals. Thus, the power of ISI at $k$th subcarrier
can be obtained by 
\begin{align*}
\mathrm{P}_{\mathrm{ISI}}\left(k\right) & =\mathrm{E\left[Y_{ISI,\mathit{k}}\cdot Y_{ISI,\mathit{k}}^{H}\right]}\\
 & =\mathbf{W}_{\mathrm{I},k}\mathrm{E\left[\mathbf{H}_{m-1}\mathbf{T}\underset{\mathbf{I}}{\underbrace{\mathbf{s}_{m-1}\mathbf{s}_{m-1}^{\mathit{H}}}}\mathbf{T}^{\mathit{H}}\mathbf{H}_{m-1}^{\mathit{H}}\right]\mathbf{W}_{\mathrm{I},\mathit{k}}^{\mathit{H}}}\\
 & =\mathbf{W}_{\mathrm{I},k}\mathrm{\underset{\mathbf{R}_{H}}{\underbrace{\mathrm{E}\left[\mathbf{H}_{m-1}\mathbf{T}\mathbf{T}^{H}\mathbf{H}_{m-1}^{H}\right]}}\mathbf{W}_{\mathrm{I},\mathit{k}}^{\mathit{H}}}
\end{align*}
where $\mathbf{W}_{\mathrm{I},k}$ denotes the $k$th row of the receive
matrix $\mathbf{W}_{\mathrm{I}}$. 

The frequency-Doppler representation of the time-varying multipath
channel is defined as $\mathbf{H}_{m}=\mathbf{F}_{2N}\tilde{\mathbf{h}}_{m}\mathbf{F}_{2N}^{H}$.
It is straightforward to show that $\left[\mathbf{H}_{m}\right]_{kn}={\displaystyle \sum_{l}}{\displaystyle \sum_{p}}\left[\tilde{\mathbf{h}}_{m}\right]_{pl}e^{j\pi\frac{nl-kp}{N}}$.
Thus, the correlation between two arbitrary elements is given by 
\begin{align*}
r_{kk'nn'}= & {\displaystyle \sum_{l,l',p,p'}}\left[\tilde{\mathbf{h}}_{m}\right]_{pl}\left[\tilde{\mathbf{h}}_{m}\right]_{p'l'}^{*}e^{j\pi\frac{nl-kp-n'l'+k'p'}{N}}\\
= & {\displaystyle \sum_{l,p,\Delta p=\Delta l}}\rho_{p-l}R_{t}\left(\Delta l\right)e^{j\pi\frac{l\left(n-n'\right)-p\left(k-k'\right)-\left(n'-k'\right)\Delta l}{N}}\\
= & \underset{R_{f}\left(k-k'\right)}{\underbrace{\left[{\displaystyle \sum_{p}\rho_{p}}e^{j\pi\frac{-p\left(k-k'\right)}{N}}\right]}}\\
 & \cdot\underset{\left[\mathbf{R}_{D}^{*}\right]_{\left(n'-k'\right)\left(n-k\right)}}{\underbrace{\left[{\displaystyle \sum_{l,\Delta l}}R_{t}\left(\Delta l\right)e^{j\pi\frac{l\left(n-n'-k-k'\right)-\left(n'-k'\right)\Delta l}{N}}\right]}}
\end{align*}
where $R_{f}\left(\cdot\right)$ is the frequency correlation function
and $\mathbf{R}_{D}$ is the Doppler covariance matrix (see Appendix
\ref{sec:Doppler-and-frequency}). As the consequence of WSSUS, the
total correlation in Doppler-frequency domain is given by the product
of frequency and Doppler correlation.

The computation of $\mathbf{R}_{\mathrm{H}}$ is as follows. The element
$r_{ij}$ of the matrix can be obtained by 
\begin{align*}
r_{ij} & =\mathrm{E\left[\mathbf{H}_{\mathit{m\mathrm{-1},i}}\mathbf{T}\mathbf{T}^{\mathit{H}}\mathbf{H}_{\mathit{m\mathit{\mathrm{-1}},j}}^{\mathit{H}}\right]}\\
 & =\mathrm{tr}\left[\mathbf{T}\mathbf{T}^{H}\mathrm{E}\left(\mathbf{H}_{\mathit{m\mathrm{-1},j}}^{\mathit{H}}\mathbf{H}_{\mathit{m\mathrm{-1},i}}\right)\right]\\
 & \mathrm{=tr\left[\mathbf{\boldsymbol{\varGamma}}_{T}\left(R_{f}\left(i-j\right)\mathbf{P}_{j}\mathbf{R}_{D}^{*}\mathbf{P}_{i}\right)\right]}\\
 & =R_{f}\left(i-j\right)\mathrm{tr}\left[\mathbf{\boldsymbol{\varGamma}}_{T}\mathbf{P}_{j}\mathbf{R}_{\mathrm{D}}^{*}\mathbf{P}_{i}\right]
\end{align*}
where $\mathbf{P}_{j}$ and $\mathbf{P}_{i}$ are permutation matrices
(moves $j$ elements down and $i$ elements to the right), $\mathbf{\boldsymbol{\varGamma}}_{T}=\mathbf{T}\mathbf{T}^{H}$,
$\mathbf{R}_{f}$ is the frequency correlation matrix determined by
the channel PDP and $\mathbf{R}_{\mathrm{D}}$ is the Doppler correlation
function determined by the time selectivity (The computation of both
correlation matrices is provided in Appendix \ref{sec:Doppler-and-frequency}
with the WSSUS assumption). Faster computation is thus row-wise (or
column-wise) 
\begin{equation}
\mathbf{r}_{j}=\mathrm{gtr\left[\mathbf{\boldsymbol{\varGamma}}_{T}\mathbf{P}_{j}\mathbf{R}_{D}^{*}\right]\circ\mathbf{R}_{f,\mathit{j}}}
\end{equation}
where $\circ$ denotes Hadamard product of two matrices, $\mathbf{r=}\mathrm{gtr}\left[\mathbf{A}\right]$
operator of square $N\times N$ matrix $\mathbf{A}$ is defined as
$r_{j}={\displaystyle \sum_{i=0}^{N-1}}a_{\left\langle i+j\right\rangle _{N},i}$.
Hence, the power of ISI can also be analytically calculated.

\section{Doppler and frequency correlation matrices\label{sec:Doppler-and-frequency}}

The frequency correlation matrix $\mathbf{R}_{f}$ is generally circulant,
the element is given by $\left[\mathbf{R}_{f}\right]_{ij}=R_{f}\left(i-j\right)$,
see \ref{eq:EDPDPFreqCorr} for EDPDP channel. It is therefore sufficient
to calculate the first row of $\mathbf{R}_{f}$ which is the Fourier
transform of the channel PDP. 

For the often ignored Doppler-domain correlation, it suffices to consider
an arbitrary $i$th tap $h_{im}$. Denote the $k$th and $p$th Doppler
response by $H_{D}\left(k\right)=\frac{1}{\sqrt{N}}{\displaystyle \sum_{m=0}^{N-1}}h_{im}e^{j\frac{2\pi}{N}mk}$
and $H_{D}\left(p\right)$, respectively. The correlation is therefore
given by 
\begin{align*}
\left[\mathbf{R}_{D}\right]_{kp} & =\frac{1}{N}\mathrm{E}\left\{ {\displaystyle \sum_{n}{\displaystyle \sum_{m}}}h_{im}h_{in}^{*}e^{j\frac{2\pi}{N}\left(mk-pn\right)}\right\} \\
 & =\frac{1}{N}\rho_{i}{\displaystyle \sum_{n}{\displaystyle \sum_{m}}}R_{t}\left(m-n\right)e^{j\frac{2\pi}{N}\left(mk-pn\right)}.
\end{align*}
Hence the correlation matrix is 2D Fourier transform of the Toeplitz
matrix with elements $\left[\mathbf{R}_{t}\right]_{nm}=R_{t}\left(m-n\right)$,
i.e., $\mathbf{R}_{D}=\mathbf{F}\mathbf{R}_{t}\mathbf{F}^{H}$.

\section{ICI and signal\label{sec:ICI-and-signal}}

Similar to the analysis of ISI, we can write 
\[
\mathbf{Y}_{\mathrm{s}}+\mathbf{Y}_{\mathrm{ICI}}=\mathbf{W}_{D}\mathbf{H}_{m}\mathbf{T}\mathbf{s}_{m}.
\]
The signal term and ICI term shall be easily separated. Consider an
arbitrary subcarrier $k$, we can obtain 
\begin{align*}
Y_{\mathrm{s},\mathit{k}} & =\mathbf{W}_{\mathrm{D},k}\mathbf{H}_{m}\mathbf{T}_{s,k}s_{m,k}\\
Y_{\mathrm{ICI},\mathit{k}} & =\mathbf{W}_{\mathrm{D},k}\mathbf{H}_{m}\bar{\mathbf{T}}_{k}\bar{\mathbf{s}}_{m,k}
\end{align*}
where $\bar{\mathbf{T}}_{s,k}$ and $\bar{\mathbf{s}}_{m,k}$ denote
the residual matrix and vector after deleting the $k$th column or
element, respectively. Thus the signal power and ICI power can be
computed according to 
\begin{align*}
\mathrm{P}_{\mathrm{S}}\left(k\right) & =\mathrm{E\left[Y_{s,\mathit{k}}\cdot Y_{s,\mathit{k}}^{H}\right]}\\
 &=\mathbf{W}_{\mathrm{D},k}\mathrm{\underset{\mathbf{R}_{H,S}}{\underbrace{\mathrm{E}\left[\mathbf{H}_{m}\mathbf{T}_{s,\mathit{k}}\mathbf{T}_{s,\mathit{k}}^{H}\mathbf{H}_{m}^{H}\right]}}\mathbf{W}_{\mathrm{D},\mathit{k}}^{\mathit{H}}}\\
\mathrm{P}_{\mathrm{ICI}}\left(k\right) & =\mathrm{E\left[Y_{ICI,\mathit{k}}\cdot Y_{ICI,\mathit{k}}^{H}\right]}\\
 &=\mathbf{W}_{\mathrm{D},k}\mathrm{\underset{\mathbf{R}_{H,ICI}}{\underbrace{\mathrm{E}\left[\mathbf{H}_{m}\bar{\mathbf{T}}_{\mathit{k}}\mathbf{\bar{T}}_{\mathit{k}}^{H}\mathbf{H}_{m}^{H}\right]}}\mathbf{W}_{\mathrm{D},\mathit{k}}^{\mathit{H}}}.
\end{align*}

\bibliographystyle{IEEEtran}
\bibliography{bibliography}

\end{document}